\documentclass[12pt]{article}
\usepackage{graphicx,epsfig,amsmath,amssymb,amsthm, mathrsfs, pstricks} 
\usepackage[left]{lineno}

\begin{document}

%\linenumbers

\title{\bf Decoherence as a sequence of entanglement swaps}

\author{{Chris Fields}\\ \\
{23 Rue des Lavandi\`{e}res, 11160 Caunes Minervois, France}\\ 
{fieldsres@gmail.com}\\
{ORCID: 0000-0002-4812-0744}}
\maketitle

\begin{abstract}
Standard semi-classical models of decoherence do not take explicit account of the classical information required to specify the system - environment boundary.  I show that this information can be represented as a finite set of reference eigenvalues that must be encoded by any observer, including any apparatus, able to distinguish the system from its environment.  When the information required for system identification is accounted for in this way, decoherence can be described as a sequence of entanglement swaps between reference and pointer components of the system and their respective environments.  Doing so removes the need for the a priori assumptions of ontic boundaries required by semi-classical models.
\end{abstract} 
~


\newpage

\section{Introduction}	

While decoherence theory is of enormous practical importance for estimating the effective lifetimes of manipulable quantum states, standard models of decoherence and methods for calculating decoherence times remain semi-classical (for reviews, see \cite{zurek:98, zurek:03, schloss:07}).  The physical mechanism of decoherence likewise remains controversial, with the environment functioning as an information sink in some formulations \cite{zeh:73, joos-zeh:85, tegmark:00, tegmark:12} and as an information channel in others \cite{zurek:04, zurek:05, zurek:06, zurek:09}.  Hence while the timecourse of decoherence can be observed experimentally \cite{brune:96, myatt:00, brune:08}, the underlying information dynamics are not yet fully characterized.

Here I show that when the observations and hence information transfer required to identify the system undergoing decoherence are taken into account, decoherence can be represented as a sequence of entanglement swaps between distinct ``reference'' and ``pointer'' components of the system and their respective environments.  The exponential decay of phase coherence for pointer observables that is derived as an approximation using semi-classical methods and observed experimentally emerges naturally in this representation.  Consistent with Landauer's principle \cite{landauer:61, landauer:99}, an ``observation'' in this setting is an exchange of energy for classical information in the form of observational outcomes, where information becomes effectively classical when it is recorded on some physical medium in a way that is thermodynamically irreversible over times much longer than the time required for the observation.  An ``observer'' in this setting is thus any system that records observational outcomes as effectively classical states.

As Krechmer \cite{krechmer:18} has emphasized, standard treatments of measurement within quantum theory, including all standard formulations of decoherence, ignore the operations required for apparatus calibration and hence ignore the calibration-relativity of observational outcomes.  The calibration process is treated here as a change in system (typically apparatus) ``settings'' that enables specific calibrating pointer-state outcomes to be obtained; see \cite{fields:18} for discussion of this approach.

\section{Reference observables and system identification}

In standard models of decoherence, the decompositional boundary between the system of interest $S$ and the decohering environment $E$, and hence the interaction $H_{SE}$ and the self-Hamiltonians $H_S$ and $H_E$ are considered to be given a priori \cite{zurek:98, zurek:03, schloss:07}.  The assumption of an $S - E$ boundary is, effectively, an assumption of classical information sufficient to specify this boundary, and hence to specify the Hilbert space $\mathcal{H}_{S}$ of $S$.  Any complete description of information flow in decoherence must account for this system-identifying information.

In practice, systems of interest are only ``given'' by observation \cite{fields:18}.  Reading the ``pointer state'' of an apparatus, for example, requires first identifying the apparatus by observing degrees of freedom other than the pointer degrees of freedom: the size, shape, layout of knobs and dials, and location of an apparatus in the laboratory are commonplace examples (Fig. 1).  These identifying or \textit{reference} degrees of freedom \cite{fields:18} must have observational outcome values, e.g. the particular shape, size, and layout of an apparatus, that are invariant over sufficiently long times to allow re-identification of the system of interest across multiple cycles of observations.  The observational outcome values of these degrees of freedom must also remain invariant during ``preparation'' operations, including calibration.  Degrees of freedom that are sufficiently stable to serve as reference degrees of freedom are generally shared by many systems other than the particular system of interest $S$; all macroscopic apparatuses, for example, have fixed sizes, shapes, and layouts.  To identify and, later, re-identify $S$, an observer must search for something having the specific, invariant values of each of these reference degrees of freedom that identify $S$: the specific shape, size, layout, etc. to be $S$ and not something else.  Finding something that satisfies all of the observational criteria to be $S$ typically requires interacting with many things besides $S$; one must, for example, typically look at many things in a cluttered laboratory to find the particular apparatus one is interested in.

\begin{figure}[ph]
\centerline{\includegraphics[width=5.0in]{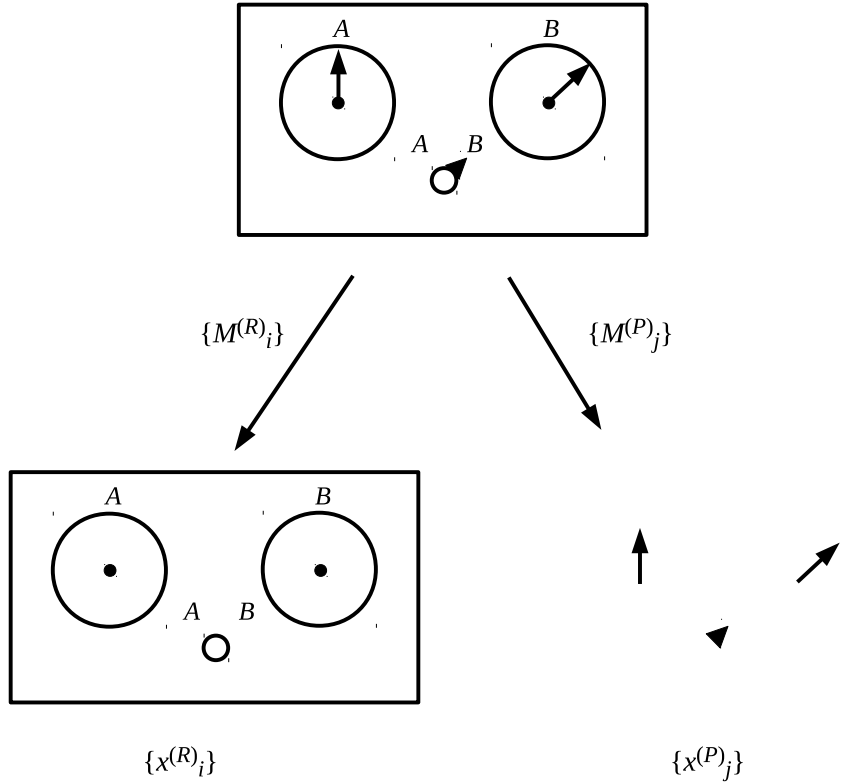}}
\vspace*{8pt}
\caption{Identifying a system of interest requires measuring a set $\{ M^{(R)}_i \}$ of reference observables with time-invariant (or only slowly-varying) outcome values $\{ x^{(R)}_i \}$, such as the particular size, shape, and layout of the system, that are distinct from the set $\{ M^{(P)}_j \}$ of pointer observables with outcomes $\{ x^{(P)}_j \}$ that indicate the system's time-varying state.  Note that the $\{ x^{(P)}_j \}$ include the values of any manipulable settings that ``prepare'' the system to be in some particular subset of pointer states.  Adapted from \cite{fields:12} Figs. 2 and 3. \label{fig1}}
\end{figure}

To make this process of searching, by observation, for the system of interest $S$ precise, consider an observer $O$ equipped with a finite collection $\{ M^{(R)}_i \}$ of Hermitian operators that act on collections of degrees of freedom within the ``world'' $W$ comprising everything other than $O$.  Assume these operators act on reference degrees of freedom as described above, i.e. on degrees of freedom that are invariant or very slowly varying over time, and can hence be considered \textit{reference observables}.  The $\{ M^{(R)}_i \}$ can, without loss of generality, be considered to pose yes/no questions and hence to have binary outcome values $\{ x^{(R)}_i \}$ in some fixed, specified basis; the observable $M^{(R)}_j$, for example, may correspond to the question ``is the color of what I am looking at black'' while $M^{(R)}_k$ may correspond to ``does it have linear dimensions $37 \times 25 \times 2.5$ cm?''  Clearly such operators can act on many different systems within $W$.  The system of interest $S$ is identified whenever all of a specified set $\{ x^{(R)}_i \}$ of reference outcome values are simultaneously obtained.  Note that in this picture, systems that yield the same set of reference outcome values when probed with the $\{ M^{(R)}_i \}$ are indistinguishable.  This inevitable \cite{moore:56} ambiguity of system identification by observers restricted to finite observational resources is decreased as the set $\{ M^{(R)}_i, x^{(R)}_i \}$ of reference observables and specified outcome values becomes large; this decrease in ambiguity is paid for by an increase in the thermodynamic resources required for system identification \cite{fields:18}.  

The reference outcome values employed to identify $S$ must, obviously, remain invariant not only during the time that $S$ is being observed, but long enough to allow re-identification of $S$ across multiple cycles of re-preparation and re-observation.  They cannot, therefore, contribute to the characterization of the time-varying state of $S$ or specify any apparatus ``settings'' or other ``preparation'' procedures, including calibration, that alter the state but not the identity of $S$.  In order to detect the time-varying state of $S$, including settings or other preparation outcomes, $O$ must also be equipped with a second finite set $\{ M^{(P)}_j \}$ of \textit{pointer observables} with binary outcome values $\{ x^{(P)}_j (t) \}$ that specify the time-varying pointer state $|P \rangle$ of $S$, including any adjustable settings or preparation outcomes.  These $\{ M^{(P)}_j \}$ must be assumed to act on whatever system is identified by the $\{ M^{(R)}_i \}$.  In practice, the number of pointer observables is typically much smaller than the number of reference observables; the number of pointers, readouts, and settings on a typical apparatus, for example, is much smaller than the number of observable properties that would need to be specified to allow a novice observer to find that apparatus in the laboratory.   Let $\{ M_k \} = \{ M^{(R)}_i \} \cup \{ M^{(P)}_j \}$ and assume that $O$ interacts with $W$ only by deploying the $M_k$ one at a time in some order.  Assuming a total of $N$ observables, we can then represent the $O - W$ interaction as:

\begin{equation} \label{HOW}
H_{OW} (t) = \sum_{k=1}^N \alpha_k (t) M_k
\end{equation}
\noindent
subject to the constraints that at all $t$,

\begin{equation}
\sum_{k=1}^N \alpha_k (t) = 1
\end{equation}
\noindent
and assuming a constant time interval $\Delta t$ to make any single observation,

\begin{equation}
\sum_{k=1}^N \int_t^{t + \Delta t} dt ~\alpha_k (t) M_k = c^{(O)} k_B T \Delta t \label{diss}
\end{equation}
\noindent
where $k_B$ is Boltzmann's constant, $T$ is temperature, and $c^{(O)} \geq$ ln2 is a measure of $O$'s thermodynamic efficiency \cite{fields:18}.  The function $\alpha_k (t)$ is naturally interpreted as the probability of deploying the measurement operator $M_k$ at $t$.  The sequence of outcomes obtained depends on the $\alpha_k (t)$; however, the incremental heat dissipation \eqref{diss} of the measurements does not.

The system of interest $S$ is, in this formulation, defined entirely implicitly: $S$ is whatever is identified by the $M^{(R)}_i$ returning the specified outcomes $x^{(R)}_i$ and its pointer state $|P \rangle$ is given by the outcomes $x^{(P)}_j$ returned by the $M^{(P)}_j$.  Note that there is no choice of measurement basis in this formulation.  The environment $E$ comprises, in this case, all degrees of freedom of $W$ except those of $S$.  With these definitions of $S$ and $E$, \eqref{HOW} embodies the assumption of an unobserved environment that justifies tracing over the state of $E$ in the environment as information sink formulation of decoherence \cite{zeh:73, joos-zeh:85, tegmark:00, tegmark:12}; the interpretation of \eqref{HOW} in the alternative environment as witness formulation is considered in \S 5 below.

If the $\{ M^{(R)}_i \}$ are to return invariant outcome values $\{ x^{(R)}_i \}$ over multiple rounds of measurement as required for system identification, they must satisfy Zurek's ``predictability sieve'' requirement.  With $H_{OW}$ given by \eqref{HOW}, this requirement can be written (cf. \cite{zurek:03} Eq. 4.41), for each operator $M^{(R)}_i$:

\begin{equation}
[H_W + H_{OW}, M^{(R)}_i] = 0 \label{commute1}
\end{equation}
\noindent
where $H_W$ is the self-Hamiltonian of $W$.  It is satisfied provided:

\begin{equation}
[M^{(R)}_i, M^{(R)}_j] = 0 \quad \mathrm{and} \quad [M^{(R)}_i, M^{(P)}_j] = 0 \label{commute2}
\end{equation}
\noindent
for all $i, j$.  Nothing, however, requires the pointer measurements $M^{(P)}_j$ to all mutually commute, as they typically do not in practice.

Making the information dynamics of system identification explicit in this way moves the physics of decoherence from the $S - E$ boundary, where it is placed in semi-classical models \cite{zurek:98, zurek:03, schloss:07}, to the $O - W$ boundary; the interaction $H_{SE}$ here remains unspecified, while $H_{OW}$ is given by \eqref{HOW}.  Decoherence is, therefore, observer-relative by definition in this representation.  As shown in the next section, it is the process of observation itself, i.e. the interaction between $O$ and $W$, not an observer-independent interaction between $S$ and a semi-classical environment, that removes coherence from $S$ in this setting.

\section{Coarse-grained measurements as entanglement swaps}

To make the information flow given by \eqref{HOW} more explicit, suppose the observer $O$ has the structure shown in Fig. 2.  Each of the operators $M_k$ is implemented by a qubit $q_k$, a classical, time-indexed outcome memory $x_k (t)$, and a single-bit recoding device that travels to the $k^{th}$ position with probability $\alpha_k$.  The rest of the observer, not represented explicitly in Fig. 2, handles energy acquisition to drive the recording process and the dissipation of waste heat.  No assumptions are made about the observer's ability to \textit{read} the memories it records; we can assume, however, that they have some effect on its behavior.

\begin{figure}[ph]
\centerline{\includegraphics[width=5.0in]{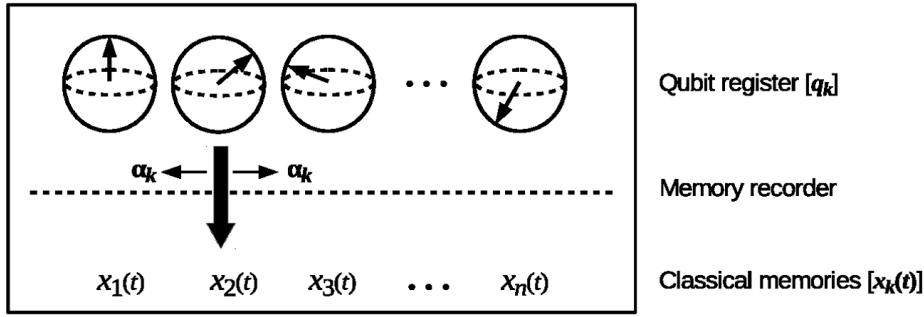}}
\vspace*{8pt}
\caption{Model observer $O$ comprising a qubit register $[q_k]$, a classical memory register $[x_k(t)]$, and a single-bit memory recording device.\protect\label{fig2}}
\end{figure}

Assume now that the system $S$ can be represented as $S = R \oplus P$, where $R$ and $P$ are the reference and pointer components, respectively, of $S$ and separability (indicated by $\oplus$) is guaranteed by \eqref{commute2}.  Consider both $R$ and $P$ to be implemented by qubit arrays $[r_i]$ and $[p_j]$ respectively, with numbers of qubits of at least the cardinality of $\{ M^{(R)}_i \}$ and $\{ M^{(P)}_j \}$ respectively.  Define the ``environments'' of $R$ and $P$, respectively, by $E_R \otimes R = E_P \otimes P = W$, where $W$ as before is everything but $O$.  Note that these definitions are consistent with $W$ being non-separable.  With these definitions, the environment $E_P$ of $P$ contains $R$, as standardly assumed when the environment of the pointer state of an apparatus, for example, is assumed to include the rest of the apparatus \cite{tegmark:00, tegmark:12}.

With these assumptions, a reference-state measurement operator $M^{(R)}_i$ monogamously entangles the observer qubit $q_i$ with the reference qubit $r_i$ and a pointer-state measurement operator $M^{(P)}_j$ monogamously entangles the observer qubit $q_j$ with the pointer qubit $p_j$.  Both operations also entail recording of the selected classical bit; completion of this recording step is the completion of the measurement operation, after which the next operation can take place.  Recording requires a time increment $\Delta t$ and dissipates $c^{(O)} k_B T$ into either $E_R$ (for reference measurements) or $E_P$ (for pointer measurements) as specified by \eqref{diss}.  Provided the recorded memory bits remain unread by $O$ or any third party, the order of measurements remains unknown, the description of the measurement process remains purely quantum mechanical, and no decoherence takes place \cite{zurek:18}.  This situation corresponds to placing the ``von Neumann cut'' somewhere outside the joint $O - S$ system and its immediately-surrounding environment, e.g. to $O$ being an isolated, unexamined apparatus into which $S$ has been embedded.

Either $O$ or some third party reading the recorded memory bits reveals the order of recording, and allows each measurement operation to be classified as either an operation on $R$ or an operation on $P$.  A human observer or an apparatus ``reporting'' an observational outcome requires such a memory-read operation.  This classification of observations by target subsystem (i.e. $R$ or $P$) effectively coarse-grains the entanglement process from the qubit scale to the multi-qubit subsystem scale.  Equivalently, third-party observations of the measurement process that distinguish operations on $R$ from operations on $P$ are coarse-grainings from the qubit scale to the subsystem scale.  Coarse-grained operations on $R$ imply monogamous $O - R$ entanglement, while coarse-grained operations on $P$ imply monogamous $O - P$ entanglement.  Alternating between system-identifying reference measurements and state-determining pointer measurements is, therefore, executing an entanglement swap between $R$ and $P$.

To see this in a simplified setting, consider $O$ to comprise just one ``observer'' qubit with state $| o \rangle$, and similarly consider $R$, $P$, $E_R$, and $E_P$ to comprise single qubits with states $| r \rangle$, $| p \rangle$, $| e_R \rangle$, and $| e_P \rangle$, respectively.  Measuring $R$, then $P$, then $R$ again corresponds, given the $R - P$ separability required by \eqref{commute2}, to the swap:

\begin{equation}
| o \otimes r \rangle \oplus | e_R \rangle \longrightarrow | o \otimes p \rangle \oplus | e_P \rangle \longrightarrow | o \otimes r \rangle \oplus | e_R \rangle   \label{swap}
\end{equation}
\noindent
where again $\oplus$ indicates separability.  This swap corresponds to an alternation of decompositional boundaries, in which $P$ and $R$ are alternately entangled with the shared ``external'' environment component $E$ comprising those degrees of freedom shared by $E_P$ and $E_R$.

\section{Exponential decay of phase coherence}

The presence of phase coherence in a system can be measured by violations of the Leggett-Garg inequality; mapping each binary outcome from $\{0, 1 \}$ to $\{ -1, 1 \}$, the Leggett-Garg inequality can be written $C_{21} + C_{32} - C_{31} \leq 1$ for $C_{ij}$ the classical correlation of measurements $i$ and $j$, and the indices referring to consecutive measurements at $t_1, t_2$ and $t_3$ \cite{emary:14}.  Sequential measurements violating the Leggett-Garg inequality are indicative of an (at least approximately) pure state of an (at least approximately) isolated system; sequential measurements satisfying the Leggett-Garg inequality are, on the other hand, indicative of mixed states and thus sampling from a classical ensemble of identically-prepared but otherwise independent and hence separable systems.  Acting with the $M^{(R)}_i$ to identify $S$ can be regarded as ``preparing'' $S$ by fixing the values of its non-pointer degrees of freedom.  Similarly, any finite set $\{ x^{(R)}_i \}$ of time-invariant reference outcome values is consistent with a classical sampling process that selects a different element of an ensemble of distinct and independent but identically-prepared systems, each characterized by the constant $\{ x^{(R)}_i \}$ \cite{moore:56}.  Evidence that a single system is being acted upon at multiple times with both the reference operators $M^{(R)}_i$ and the pointer operators $M^{(P)}_j$ is only obtained if the pointer outcomes $x^{(P)}_j$ obtained exhibit Leggett-Garg inequality violations \cite{fields:18}. 

Suppose now an ensemble of systems $< S_i > = < R_i \oplus P_i>$ that are indistinguishable by the reference operators $ M^{(R)}_i$, and for each of which a single, fixed pointer operator $M^{(P)}$ acts on a single qubit of each $P_k$ to yield an outcome $x^{(P)}$ and project a state $|p_k \rangle$.  Suppose further that the $M^{(R)}_i$ are all executed once, in a fixed sequence, followed by $M^{(P)}$ in each cycle of measurements, with each operation requiring a time interval $\Delta t$ as before.  The standard ``picture'' of measurement in which some particular system $S = S_k$ is given a priori, and hence system identification by the $M^{(R)}_i$ is unnecessary, corresponds in this case to a sequence of executions of $M^{(P)}$:

\begin{equation*}
M^{(P)} (t) \dots M^{(P)} (t + \Delta t) \dots M^{(P)} (t + 2 \Delta t) \dots M^{(P)} (t + 3 \Delta t) \dots
\end{equation*} 
\noindent
Here the projection postulate implies that each action of $M^{(P)}$ re-prepares the state $| p_k \rangle$ of the single pointer qubit of the given $S_k$; hence if $\Delta t$ is small compared to the timescale of interactions between $P_k$ and its environment $E_{P_k}$, all executions of $M^{(P)}$ return the same outcome value $x^{(P)}$.  This is the familiar quantum Zeno effect \cite{misra:77}.  Note that it assumes that repeated interactions with the observer, i.e. executions of $M^{(P)}$, maintain phase coherence and so do not decohere $P_k$.  This corresponds to the swaps to $| r \rangle$ being removed from \eqref{swap}, leaving the fixed state $| o \otimes p_k \rangle \oplus | e_{P_k} \rangle$.

If, however, $n - 1$ reference operators $M^{(R)}_i$ are executed between each execution of $M^{(P)}$, the delay between executions of $M^{(P)}$ is increased:

\begin{equation*}
M^{(P)} (t) \dots M^{(P)} (t + n \Delta t) \dots M^{(P)} (t + 2n \Delta t) \dots M^{(P)} (t + 3n \Delta t) \dots
\end{equation*} 
\noindent
This increased delay between pointer measurements allows more time for $P_k - E_{P_k}$ interactions.  Coupling a degree of freedom $\xi$ of $E_{P_k}$ to $P_k$ is, effectively, exchanging the pointer component $P_k$ for the pointer component $P_k \otimes \xi$.  Provided the $M^{(R)}_i$ do not (detectably) act on $\xi$, the commutativity conditions \eqref{commute2} remain satisfied and this new pointer component $P_k \otimes \xi$ remains the pointer component of some element of the ensemble $< S_i >$ of systems that are indistinguishable by the $M^{(R)}_i$.  If the probability of such a coupling to a degree of freedom of $E_{P_k}$ is $P_{int} > 0$ in any interval $\Delta t$ during which a reference operator is executed, the probability that the same pointer component $P_k$ and hence the same single element $S_k$ of $< S_i >$ is being probed on every cycle of measurement is, after $m$ cycles:

\begin{equation}
Prob (\mathrm{pure}) = (1 - P_{int})^{m(n - 1)} \longrightarrow 0 ~\mathrm{for} ~m >> 1 ~\mathrm{or} ~n >> 1 \label{prob}
\end{equation}
\noindent
In semi-classical models of decoherence as resulting from otherwise-unobserved scattering of environmental particles, this exchange of $P_k$ for $P_k \otimes \xi$ is implemented by scattering events, with $n - 1$ corresponding to the particle number density in the scattering constant and $mn \Delta t$ to the total elapsed time \cite{schloss:07}. 

The practical implication of \eqref{prob} can be seen by considering Schr\"{o}dinger's cat.  If $n = 1$, there are no swaps to $R$ and the cat state $(|\mathrm{dead} \rangle + |\mathrm{alive} \rangle) / \sqrt{2}$ can be maintained indefinitely by the quantum Zeno effect.  In this case, however, no reference degrees of freedom are ever measured, so nothing identifies $(|\mathrm{dead} \rangle + |\mathrm{alive} \rangle) / \sqrt{2}$ as the state of a \textit{cat}.  Introducing reference operators $M^{(R)}_i$ to identify the cat introduces swaps to $R$ and the probability of operating repeatedly on a pure state of a single element of the ensemble $< S_i >$ of cat-like entities indistinguishable by the $M^{(R)}_i$ exponentially decreases as specified by \eqref{prob}, i.e. the cat state decoheres.  These reference operators can act on whatever is singled out to function as a reference component $R$; identifying the exterior of the box and measuring heat radiated from it equally decoheres $(|\mathrm{dead} \rangle + |\mathrm{alive} \rangle) / \sqrt{2}$ \cite{nielsen:00}.  What \eqref{prob} tell us, therefore, is that state purity can be maintained only if system identification is abandoned or at least minimized (i.e. $n$ is kept very small).  The states of strongly-identified systems - e.g. macroscopic objects with well-defined sizes, shapes, and positions for which $n >> 1$ - decohere almost immediately.

\section{Einselection and the $S - E$ boundary}

The complete set of observational outcomes $\{ x_k \} = \{ x^{(R)}_i \} \cup \{ x^{(P)}_j \}$ is also the set of eigenvalues of the interaction $H_{OW}$ between the observer and the observed world defined by \eqref{HOW}.  These eigenvalues encode, by assumption, all of the information that $O$ can obtain about $W$.  This set can clearly be made arbitrarily large by including reference observables and sets of specified outcome values that identify and pointer observables that characterize many different systems; such an increase in observational capability must, as noted earlier, be paid for by an increase in thermodynamic resources.

In the ``environment as witness'' formulation of decoherence, the external environment $E$ serves as an information channel from $S$ to $O$ \cite{zurek:04, zurek:05}; this formulation is extended to multiple, non-interacting observers in quantum Darwinism \cite{zurek:06, zurek:09}.  The information transmitted by $E$ in these formulations is an encoding of the eigenvalues of the interaction $H_{SE}$, an interaction defined at the $S - E$ boundary.  As noted earlier, this boundary and hence $H_{SE}$ are taken as given a priori.  Decoherence is implemented at this boundary by the predictability sieve requirement \eqref{commute1}, a process termed ``environment-induced superselection'' or einselection \cite{zurek:98}.

If $E$ is treated as an observer and hence identified with $O$ and $S$ is identified with $W$, einselection is decoherence as described by \eqref{HOW}, \eqref{commute1}, and \eqref{commute2}.  The invariant eigenvalues $x^{(R)}_i$ in this case specify the $S - E$ boundary to $E$ and the pointer eigenvalues $x^{(P)}_j$ are the information einselected on that boundary.  This remains true when an external observer is considered and both $S$ and $E$ become components of $W$ as above (Fig. 3).

\begin{figure}[h]
\centerline{\includegraphics[width=6.0in]{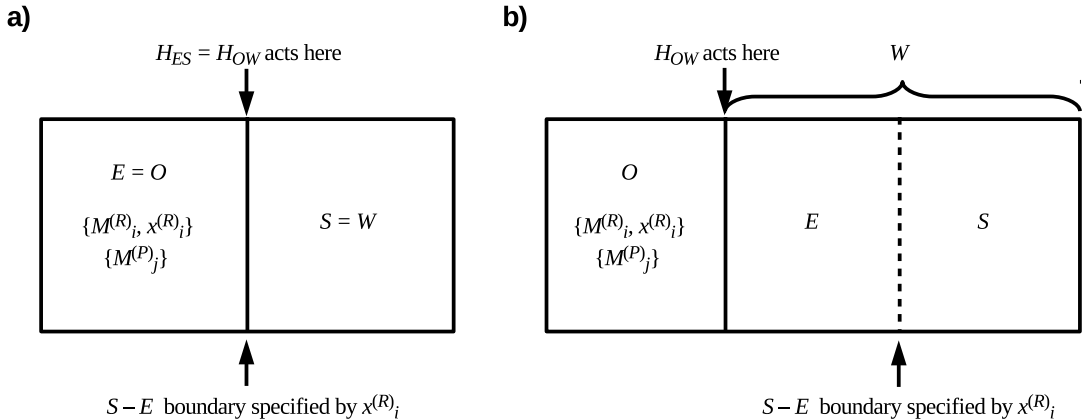}}
\vspace*{8pt}
\caption{(a) If the external environment $E$ is treated as an observer, einselection is decoherence at the $S - E$ boundary.  (b) Interposing $E$ as a channel between $O$ and $S$ changes neither the definition of $H_{OW}$ nor the specification of the $S - E$ boundary by the reference eigenvalues $x^{(R)}_i$. \label{fig3}}
\end{figure}

The present, fully observation-based formulation of decoherence thus clarifies two points of ambiguity present in traditional, environment as information sink formulations but made more obvious in the environment as witness formulation.  First, it shows that the classical information specifying the location of the $S - E$ boundary is encoded by $O$, however $O$ is defined.  This information is encoded, specifically, by the invariant $S$-identifying eigenvalues $x^{(R)}_i$.  This answers Zurek's question of ``how one can define systems given an overall Hilbert space `of everything' and the total Hamiltonian'' (\cite{zurek:98} p. 1794) without the need to postulate objectively-bounded, ontic systems in contravention of the associativity of Hilbert-space decomposition \cite{zanardi:01, zanardi:04, dugic:06, dugic:08}.  Second, it shows that the encoding redundancy required by quantum Darwinism is implemented by redundant encoding of the $x^{(R)}_i$ by multiple observers.  There is no need to assume an objective, ontic redundancy of encoding by the environment \cite{fields:10}.

\section{Conclusion}

Standard semi-classical models of decoherence do not take explicit account of the classical information required to specify the system - environment boundary.  It is shown here that when this is done, decoherence can be described as a sequence of entanglement swaps between reference and pointer components of the system and their respective environments.  The classical information specifying the boundary is encoded by the set $\{ x^{(R)}_i \}$ of reference eigenvalues that must be encoded in memory by any observer, including any apparatus, able to distinguish the system of interest $S$ from the external environment $E$.  

This formulation of decoherence renders both the $S - E$ boundary and the decoherence process itself observer-relative.  It thus represents the physics of decoherence in a way that is broadly consistent with the relational view of quantum theory introduced by Rovelli \cite{rovelli:96}.  In this picture, $S - E$ interactions that induce decoherence, e.g. scattering by environmental particles in semi-classical models, are equivalent to observer-relative exchanges between observationally-indistinguishable systems \cite{fields:12b}.  ``Systems'' in this picture are, therefore, not ontic but rather FAPP \cite{bell:90}.

\section*{Acknowledgments}

This research was funded by the Federico and Elvia Faggin Foundation.  Thanks to participants in the Quantum Contextuality in Quantum Mechanics and Beyond 2018 workshop for relevant discussions and to an anonymous referee for helpful comments.

\end{document}